\documentclass{article}

\usepackage{spconf}
\usepackage[utf8]{inputenc}
\usepackage{amsmath}
\usepackage{mathtools}
\usepackage{bm}
\usepackage{graphicx}
\usepackage{amsfonts}
\usepackage{url}

\def\prec{\mathbf{D}}
\def\lap{\mathbf{L}}
\def\adj{\mathbf{A}}
\def\deg{\mathbf{B}}

\newcommand{\iid}{\stackrel{iid}{\sim}}

\title{Anatomically informed Bayesian spatial priors for fMRI analysis}
\name{David Abramian$^{\star \dagger}$, Per Sid{\'e}n$^{\ddagger}$, Hans Knutsson$^{\star \dagger}$, Mattias Villani$^{\ddagger \uparrow}$, Anders Eklund$^{\star \dagger \ddagger}$
\thanks{This work was supported by the Swedish Research Council, grant 2017- 04889, and by the Center for Industrial Information Technology (CENIIT) at Linköping University }
}

\address{$^{\star}$ Division of Medical Informatics, Department of Biomedical Engineering\\ $^{\dagger}$ Center for Medical Image Science and Visualization (CMIV) \\   $^{\ddagger}$ Division of Statistics \& Machine learning, Department of Computer and Information Science \\ Linköping University, Linköping, Sweden \\ $^{\uparrow}$ Department of Statistics, Stockholm University, Stockholm, Sweden}

\begin{document}
%
\maketitle

\begin{abstract}
Existing Bayesian spatial priors for functional magnetic resonance imaging (fMRI) data correspond to stationary isotropic smoothing filters that may oversmooth at anatomical boundaries. We propose two anatomically informed Bayesian spatial models for fMRI data with local smoothing in each voxel based on a tensor field estimated from a $T_1$-weighted anatomical image. We show that our anatomically informed Bayesian spatial models results in posterior probability maps that follow the anatomical structure.
\end{abstract}

\begin{keywords}
    Bayesian statistics, functional MRI, activation mapping, adaptive smoothing
\end{keywords}

\vspace{-0.3cm}
\section{Introduction}
\vspace{-0.1cm}
\label{sec:intro}
The analysis of functional magnetic resonance imaging (fMRI) data has generally relied on frequentist statistics to perform inferences about brain activity. After fitting a general linear model (GLM) to the individual voxel time series, t-values are calculated for contrasts of interest at every voxel, resulting in t-maps that can be thresholded at an appropriate significance level. Isotropic Gaussian smoothing is the most common way of preprocessing fMRI data, but several adaptive smoothing approaches for detecting brain activity have been proposed~\cite{friman2003adaptive, nandy2003novel, rydell2008bilateral, eklund2011fast, behjat2015anatomically, zhuang2017family, lohmann2018lisa}.

During the 2000s, a Bayesian framework for fMRI analysis was developed by Penny et al.~\cite{penny2003variational, penny2005bayesian, penny2007bayesian,harrison2008diffusion,harrison2008graph}, which provided increased flexibility compared to the classical frequentist approach, by allowing the estimation of individual smoothness parameters for each regressor and autoregressive (AR) noise coefficient. Sidén et al.~\cite{siden2017fast,siden2019} extended this work to an efficient Markov Chain Monte Carlo (MCMC) implementation which works in 2D as well as 3D. However, we are not aware of any work that performs anatomically adaptive Bayesian spatial modeling. In this work we propose two approaches for performing 2D Bayesian spatial modeling in an anatomically-adaptive way.

\section{Methods}

\subsection{Bayesian spatial GLM}
The most common way to model fMRI data is as a voxel-wise general linear model with serial correlations in the residuals, which are modeled as an AR($p$) process. This combined model is referred to as GLM-AR($p$). In the case of single subject data, with $T$ volumes, $N$ voxels, $K$ regressors and an AR model of order $p$, this model is written as
\begin{equation*}
    \underset{[T \times N]}{\mathbf{Y}} = \underset{[T \times K]}{\mathbf{X}}\underset{[K \times N]}{\mathbf{W}} + \underset{[T \times N]}{\mathbf{E}} ,
\end{equation*}
\begin{equation*}
    \underset{[T \times N]}{\mathbf{E}} = \underset{[T \times P]}{\widetilde{\mathbf{E}}}\underset{[P \times N]}{\mathbf{R}} + \underset{[T \times N]}{\mathbf{Z}} ,
\end{equation*}
where $\mathbf{Y}$ is the observation matrix, $\mathbf{X}$ is the design matrix, $\mathbf{W}$ is the regressor matrix and $\mathbf{E}$ is the residual matrix. The residual matrix $\mathbf{E}$ is itself expressed as the product of a lagged prediction error matrix $\widetilde{\mathbf{E}}$  with an AR coefficient matrix $\mathbf{R}$ plus a matrix of i.i.d. zero mean Gaussian errors $\mathbf{Z}$, where $\mathbf{Z}_{\cdot,n} \iid N(\mathbf{0}, \lambda_n^{-1} \mathbf{I}_T)$,  and $\lambda_n$ is the noise precision at voxel $n$. In this Bayesian framework, smoothness of the regression coefficients enters the model through a spatial precision matrix $\prec_w$ in their prior distribution, according to 
\begin{equation*}
    \mathbf{W}_{k,\cdot}' \sim \mathcal{N} \left( \mathbf{0}, \alpha_{k}^{-1} \prec_w^{-1} \right) ,
\end{equation*}
where $\mathbf{W}'_{k,\cdot}$ is the transpose of the $k$-th row of the regression coefficient matrix $\mathbf{W}$, representing the $k$-th regression coefficient at every voxel, and $\alpha_k$ is a smoothness hyperparameter for the $k$-th regressor (since the following also applies to the AR coefficients, we refer to their precision matrices generically as $\mathbf{D}$). The precision matrix $\mathbf{D}$ encodes the conditional dependencies between every pair of voxels. The smoothness assumption constrains these dependencies to exist only between a voxel and its immediate neighbors, which has the advantage of making this matrix very sparse. 

\subsection{Uniform graph Laplacian precision matrix}
Graphs constitute a natural way of describing relationships between sets of elements. Therefore, the precision matrices employed by Penny et al. and Sidén et al. take the form of graph Laplacian matrices. In their formulation the precision matrix $\prec$ is an unweighted graph Laplacian (UGL), where $d_{i,i}$ equals the number of pixels neighboring pixel $i$ and ${d_{i,j}=-1}$ if pixels $i,j$ are cardinal neighbors. For inner pixels this corresponds to the Laplacian operator, which in 2D takes the form
\small
\begin{equation*}
    \mathbf{H}_{\text{UGL}} = \begin{pmatrix*}[r]
        0 & -1 & 0  \\ 
        -1 & 4 & -1 \\
        0 & -1 & 0
    \end{pmatrix*} .
\end{equation*}
\normalsize

Generically, graph Laplacian matrices can be constructed with~\cite{chung1997spectral}
\begin{equation*}
    \label{eq:laplacian}
    \lap = \deg - \adj ,
\end{equation*}
where $\adj$ is an \emph{adjacency matrix} whose $a_{i,j}$ element, called a \emph{weight}, represents in this context the strength of the conditional dependence between pixels $i$ and $j$, and $\deg$ is a diagonal \emph{degree matrix} with $b_{i,i} = \sum_j a_{i,j}$, that is, the sum of all the incoming weights to pixel $i$. Since $\deg$ can be constructed from $\adj$, the adjacency matrix is sufficient for generating the graph Laplacian. Under this formulation the same UGL prior can be constructed by considering its dependency structure (neighborhood), which for inner pixels takes the shape 
\small
\begin{equation*}
    \mathbf{N}_{\text{UGL}} = \begin{pmatrix*}[r]
        0 & 1 & 0  \\ 
        1 & 0 & 1 \\
        0 & 1 & 0
    \end{pmatrix*} ,
\end{equation*}
\normalsize
that is, $a_{i,j}=1$ if pixels $i,j$ are cardinal neighbors. It can be seen that this results in the same precision matrix $\prec \equiv \lap$.

Due to the uniform neighborhood shape and constant weights between neighbors this type of prior is incapable of encoding anatomical information. Such a prior will not respect anatomical tissue boundaries, mixing, for example, signals from white and gray brain matter. Conversely, in order for the prior to encode relevant anatomical information it is necessary for the dependencies between neighboring pixels to be specified independently at each location.

\subsection{The structure tensor}
In order to establish pixel relationships that respect anatomical boundaries, we must first estimate the position and orientation of these boundaries. Our main tool for achieving this is the structure tensor, a tensor estimated at every pixel (using quadrature filters along different directions~\cite{granlund1995signal, knutsson2003what}) whose eigenvalues and eigenvectors reveal the degree to which spatial structure is present and the orientation of any such structure, here understood as lines or edges. A detailed explanation of the estimation and interpretation of structure tensors falls outside the scope of this paper, see~\cite{knutsson1989representing, granlund1995signal, knutsson2011representing} for a thorough treatment of this topic. For our purposes it is sufficient to point out that the eigenvector of the structure tensor corresponding to the largest eigenvalue is perpendicular to the main structure orientation at the given pixel, while the eigenvalue is proportional to the degree of certainty of this orientation. The second eigenvalue indicates the amount of structure in the orientation perpendicular to the first eigenvector. The structure tensor is, therefore, a flexible model capable or representing structure in zero (noise, uniform data), one (line, edge), or two (crossing lines, crossing edges) orientations (for the $2$D case).

\begin{figure}[tbp]
    \centering
    \includegraphics[width=0.85\columnwidth]{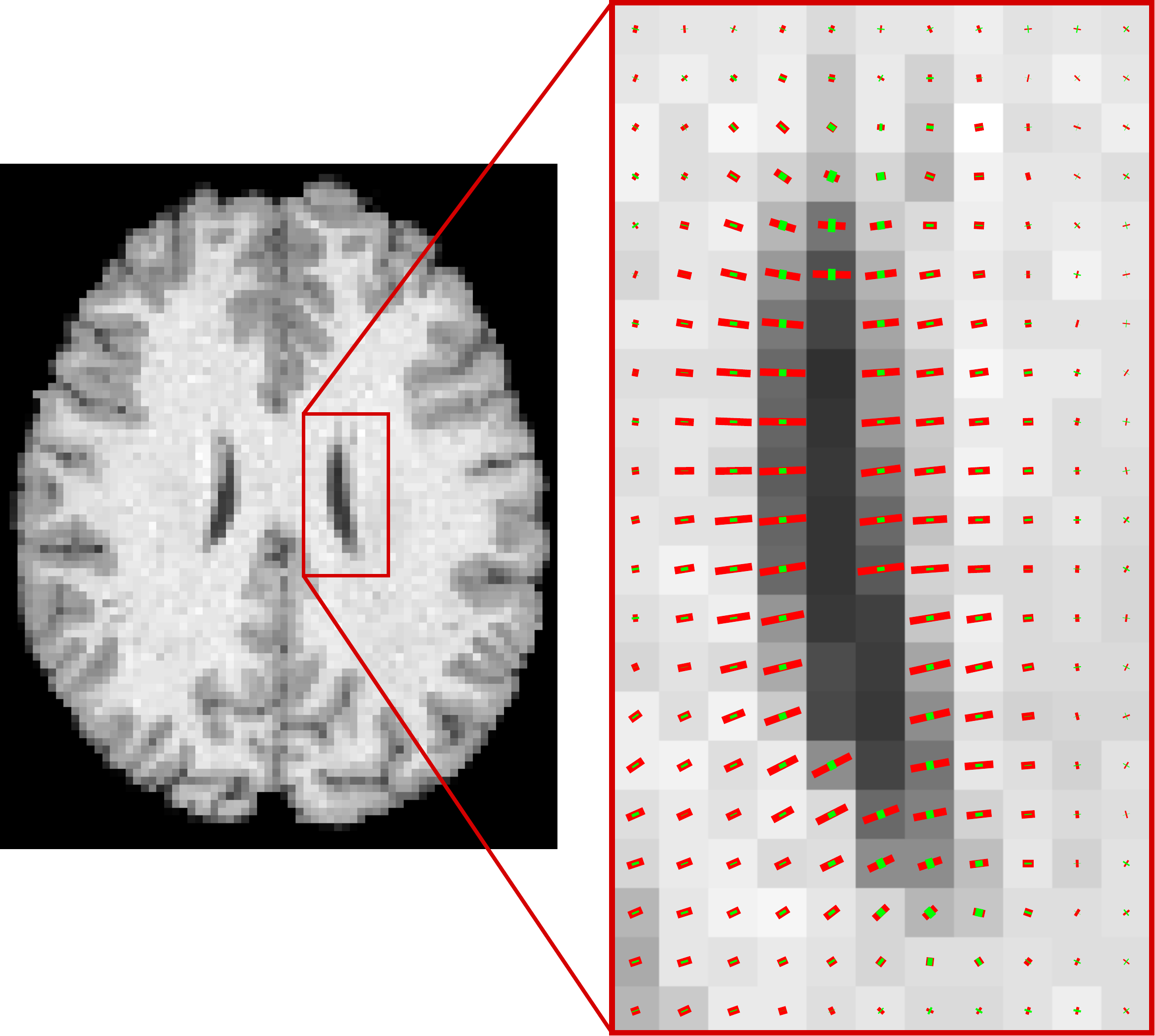}
    \caption{Local orientation represented using the structure tensor. Red vectors indicate the main local orientation, while green vectors indicate the second (perpendicular) orientation. If no vector is present, there is no orientation information available (e.g. due to uniform intensity).}
    \label{fig:struc_tens}
\end{figure}

As fMRI data lacks contrast, the structure tensor is calculated from a registered $T_1$-weighted volume. Figure~\ref{fig:struc_tens} shows the eigenvectors of the structure tensor for a typical brain image. The tensor is small and isotropic in very uniform regions, while next to lines or edges it becomes anisotropic and oriented perpendicularly to the line or edge.

\subsection{Spatial model with four orientations}
In our first proposed solution we assign to every pixel one of four oriented neighborhood structures according to the local structural orientation in that pixel. Such a model offers limited angular resolution, since only four different orientations can be represented. However, it simplifies the formulation of the possible neighborhoods and provides increased sparsity in the precision matrix $\prec$. We refer to this model as 4DIR. We start by considering four possible orientations: horizontal, vertical, and two diagonals. The corresponding orientation vectors are
\small
\begin{equation*}
    \mathbf{d}_x = 
    \begin{pmatrix}
        1 \\ 
        0
    \end{pmatrix}, \quad   
    \mathbf{d}_y = 
    \begin{pmatrix}
        0 \\ 
        1
    \end{pmatrix},   
\end{equation*}
\begin{equation*}
    \mathbf{d}_{xy} = \frac{1}{\sqrt{2}} 
    \begin{pmatrix}
        1 \\ 
        1
    \end{pmatrix}, \quad
    \mathbf{d}_{-xy} = \frac{1}{\sqrt{2}} 
    \begin{pmatrix}
        -1 \\ 
        1
    \end{pmatrix}.
\end{equation*}
\normalsize

For computational efficiency we want to avoid having to calculate the eigenvectors of the structure tensor. We can find which of the four orientations is closest to the main tensor orientation at each point by first defining tensors corresponding to each of the four orientations
\small
\begin{equation*}
    \mathbf{T}_x = \mathbf{d}_x \mathbf{d}_x^T, \quad
    \mathbf{T}_y = \mathbf{d}_y \mathbf{d}_y^T, 
\end{equation*}
\begin{equation*}
    \mathbf{T}_{xy} = \mathbf{d}_{xy} \mathbf{d}_{xy}^T, \quad
    \mathbf{T}_{-xy} = \mathbf{d}_{-xy} \mathbf{d}_{-xy}^T,
\end{equation*}
\normalsize
and then projecting, through an inner product, the structure tensor at each point onto the four orientation tensors. The maximum projection value will correspond to the orientation closest to that of the structure tensors.

Having found at each pixel which of the four orientations is closest to that of the structure tensor, we define neighborhoods (filters) along each of the four orientations
\small
\begin{equation*}
    \mathbf{N}_x = 
    \begin{pmatrix*}[r]
        0 & 0 & 0 \\ 
        1 & 0 & 1 \\
        0 & 0 & 0
    \end{pmatrix*},
    \quad
    \mathbf{N}_y = 
    \begin{pmatrix*}[r]
        0 & 1 & 0 \\ 
        0 & 0 & 0 \\
        0 & 1 & 0
    \end{pmatrix*},
\end{equation*}
\begin{equation*}
    \mathbf{N}_{xy} = 
    \begin{pmatrix*}[r]
        1 & 0 & 0 \\ 
        0 & 0 & 0 \\
        0 & 0 & 1
    \end{pmatrix*},
    \quad
    \mathbf{N}_{-xy} = 
    \begin{pmatrix*}[r]
        0 & 0 & 1 \\ 
        0 & 0 & 0 \\
        1 & 0 & 0
    \end{pmatrix*}.
\end{equation*}
\normalsize
As the structure tensor is aligned across the orientation of lines and edges (see Figure~\ref{fig:struc_tens}), which we are trying to preserve, we associate to each pixel a neighborhood perpendicular to the orientation of the structure at that point (e.g. if $\mathbf{d}_x$ is the orientation for some pixel, then the appropriate neighborhood would be $\mathbf{N}_y$).

Having assigned a specific neighborhood to each pixel, a new adjacency matrix $\adj_{\text{4DIR}}$ can be constructed and used to define a graph Laplacian $\prec_{\text{4DIR}}$. However, due to the specific process for calculating the orientation at each position, it cannot be guaranteed that the relationship between a pair of pixels $i,j$ is symmetric, i.e., that $a_{i,j}=a_{j,i}$, which would preclude $\prec_{\text{4DIR}}$ from being used as a precision matrix. Symmetry can be enforced by applying the procedure
\begin{equation*}
    \mathbf{A_{\text{4DIR}}}' = \frac{\mathbf{A_{\text{4DIR}}}+\mathbf{A_{\text{4DIR}}}^{T}}{2} ,
\end{equation*}
which effectively sets the value of $a_{i,j}$ and $a_{j,i}$ to the average of the two. This correction has little effect on the orientations encoded in the graph and allows the new matrix $\lap_{\text{4DIR}}'$ to be used as a precision matrix $\prec_{\text{4DIR}}$.

\subsection{Spatial model with arbitrary orientations}
\begin{figure}[tbp]
    \centering    
    \includegraphics[width=0.6\columnwidth]{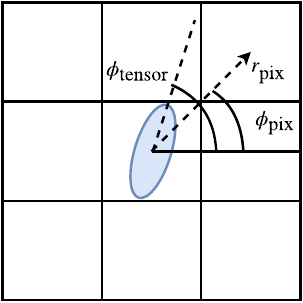}
    \caption{Variables involved in weighting of the neighborhoods for the ANYDIR model. The structure tensor is shown in the middle.}
    \label{fig:ANYDYR_illustration}
\end{figure}

The previously presented model is limited in angular resolution, as it can only encode four orientations. In order to overcome this limitation we consider a $3 \times 3$ neighborhood around every pixel and determine the weight for each of the $8$ possible neighbors by sampling from a continuous function at discrete positions corresponding to the centers of the neighboring pixels. We refer to this model as ANYDIR. The sampled function is
\begin{equation*} \label{eq:better}
    a_{i,j} = \mathbf{f}(i,j) = \frac{\left| \sin(\phi_{\mathrm{pix}_j} - \phi_{\mathrm{tensor}_i}) \right| ^\alpha}{r_{\mathrm{pix}_j}^\beta} , \quad \alpha, \beta > 0,
\end{equation*}
where $\phi_{\mathrm{pix}_j}$ is the angle of the line connecting the central pixel $i$ with neighboring pixel $j$, $\phi_{\mathrm{tensor}_i}$ is an angle representing the main orientation of the structure tensor at pixel $i$, $r_{\mathrm{pix}_j}$ is the distance between pixels $i$ and $j$ (see Figure~\ref{fig:ANYDYR_illustration}). Additionally, $\alpha$ and $\beta$ are adjustable parameters taking non-negative real values, where $\alpha$  controls the width of the distribution around $\phi_{\mathrm{tensor}_i}$ and $\beta$ penalizes the values in the diagonal neighbors with respect to the horizontal and vertical ones. In this work we use $\alpha=12$ and $\beta=5$.

This neighborhood formulation generates neighborhoods that vary continuously with respect to $\phi_{\mathrm{tensor}_i}$, which allows the representation of arbitrary orientations in the structures around pixels, at the cost of reduced sparsity in the precision matrix. As in the previous case the adjacency matrix $\adj_{\text{ANYDIR}}$ is not guaranteed to be symmetric. The same kind of correction can be applied in order to generate a graph Laplacian matrix $\lap'_{\text{ANYDIR}}$ suitable for use as a precision matrix $\prec_{\text{ANYDIR}}$.

\section{Results}
We compared the results produced by all three precision matrices in the context of Bayesian fMRI analysis. We used preprocessed data from subject $100307$ of the Human Connectome Project~\cite{van2013wu}, specifically the $T_1$-weighted volume and the motor task data. The analysis was performed slice-by-slice using the SPM package for Matlab together with the extension developed by Sidén et al.~\cite{siden2017fast}. The only modification to the code was the addition of the proposed precision matrices. Our modified code can be found at (\url{https://github.com/DavidAbramian/adaptiveBayesianPrior}). 

As a preprocessing step, the $T_1$-weighted volume was downsampled to match the resolution of the fMRI data. This is necessary, since the precision matrix has to be estimated from the $T_1$-weighted volume and later applied to the fMRI data.

Due to the time-consuming nature of the MCMC analysis it was carried out on a single representative axial slice. The Gibbs sampling algorithm was iterated $10,000$ times, with $1,000$ warmup iterations, and with a thinning factor of $5$.

\subsection{Anatomical adaptiveness}
\begin{figure}
    \centering
    \includegraphics[width=0.9\columnwidth]{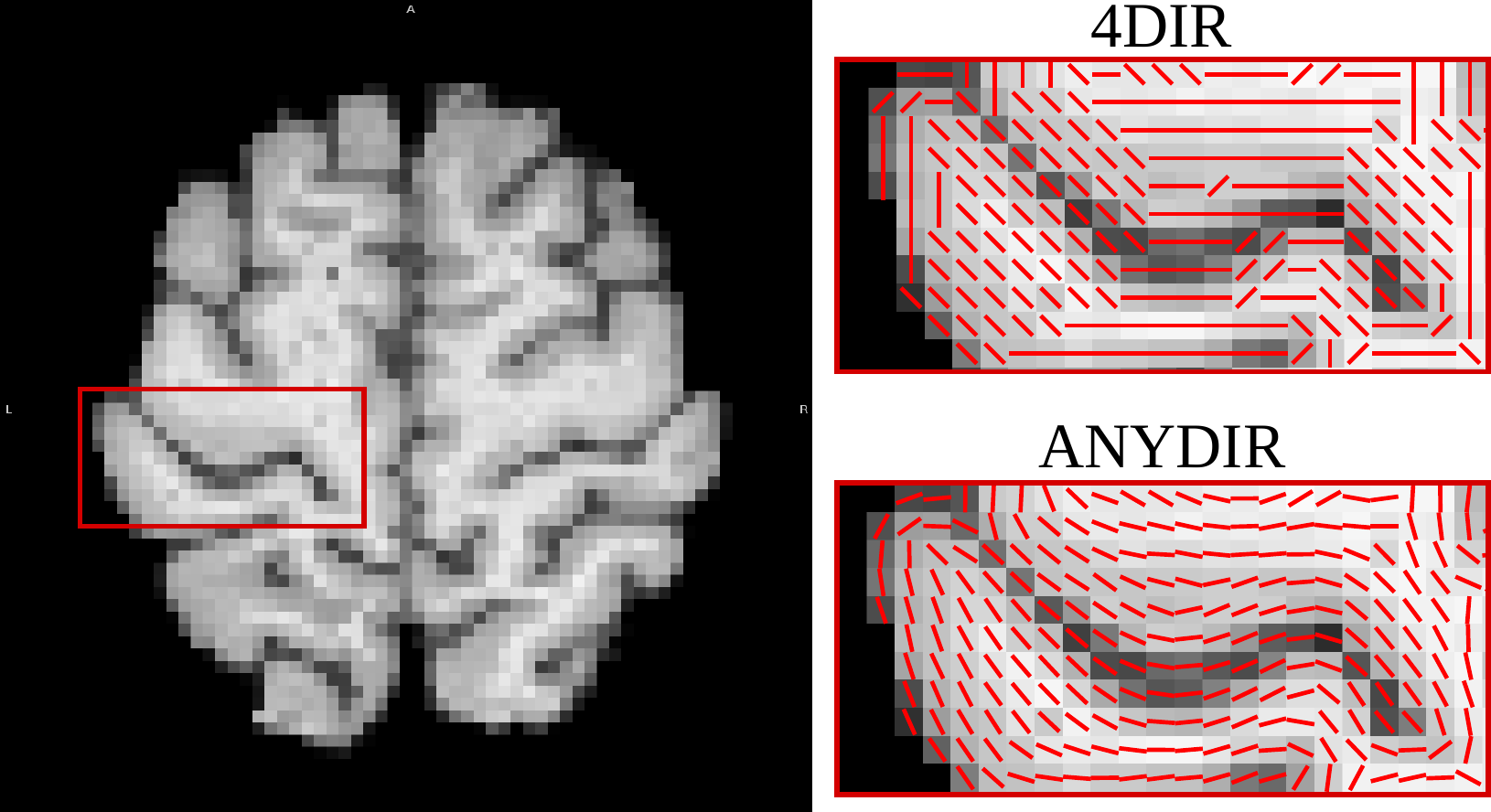}
    \caption{Comparison of neighborhood structures implied by both of the proposed models. Lines represent the orientation of spatial prior dependencies at the given point. Both methods adapt these dependencies at each point in accordance with the anatomical structure. The 4DIR method only makes use of four possible orientations, while with ANYDIR the orientations vary continuously.}
    \label{fig:comparison}
\end{figure}

Figure~\ref{fig:comparison} illustrates the orientation of the pixel neighborhoods generated by the two proposed approaches, which determine the conditional dependencies encoded in prior precision matrix $\prec$. In both cases the models have successfully adapted to the anatomical structure given by the $T_1$-weighted volume, as spatial prior dependence is placed along lines and edges and not across them. The 4DIR method shows limited angular resolution, as it can only represent neighborhoods in four orientations, while the ANYDIR produces neighborhoods in arbitrary orientations.

\subsection{Functional MRI results}
\begin{figure}[tb]
    \centering
    \includegraphics[width=\columnwidth]{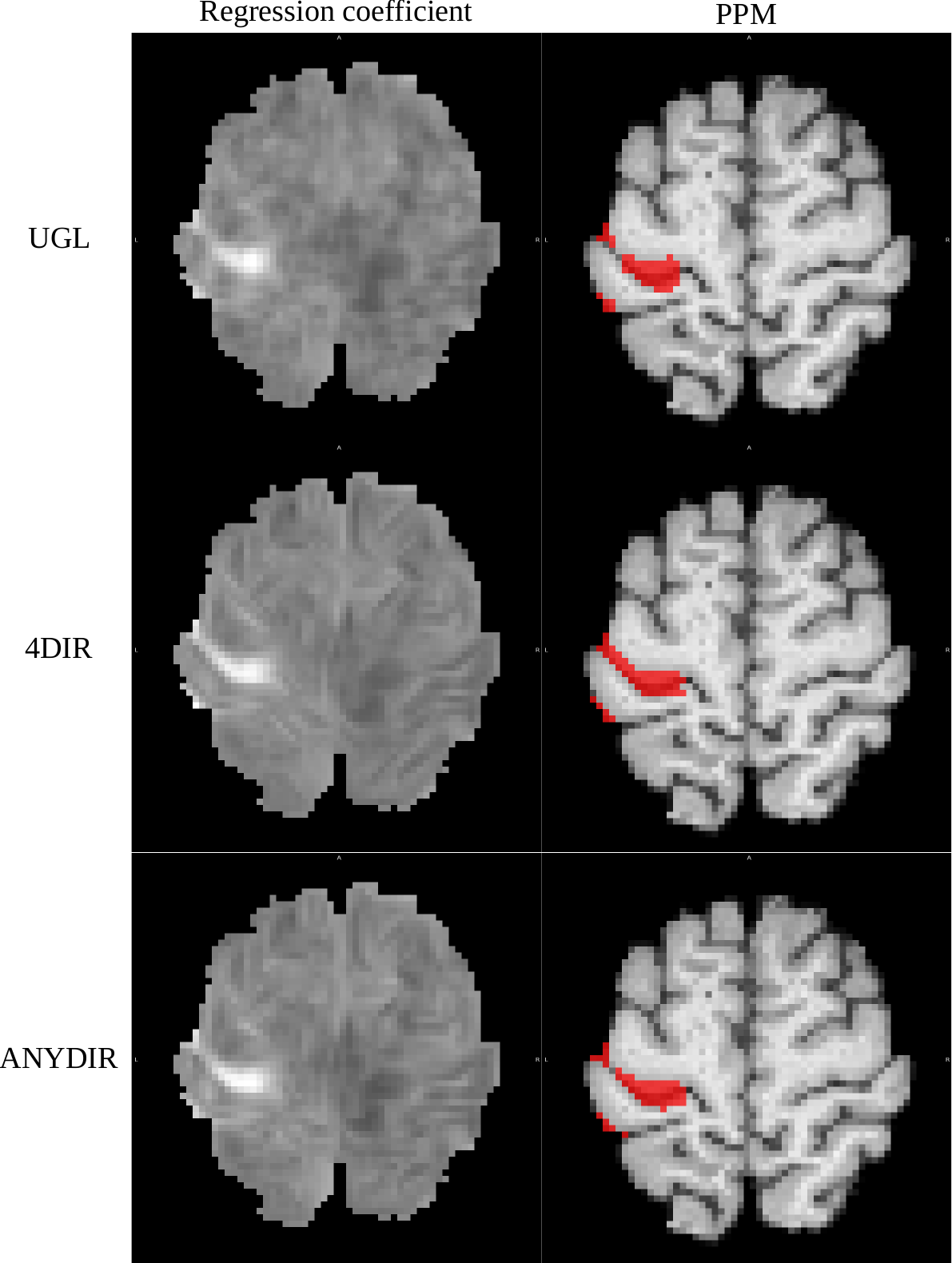}
    \caption{Bayesian GLM regression results obtained using MCMC. Left: regression coefficient corresponding to right hand motor task. Right: PPMs for the probability of effect of the right hand motor task exceeding $0.2\%$ of the global mean signal, thresholded at $0.8$. Top: UGL model. Middle: 4DIR model. Bottom: ANYDIR model.}
    \label{fig:results}
\end{figure}

Figure~\ref{fig:results} shows, for all three models, the regression coefficients obtained for a right hand motor task, as well as posterior probability maps (PPMs) quantifying the probability of the effect of said task exceeding $0.2\%$ of the global mean signal and thresholded at $0.8$. The regression coefficients for both of the proposed models clearly reflect anatomical spatial patterns absent from the UGL results. The patterns are highly angular for the 4DIR model as a result of the limited angular resolution, while the ANYDIR model results in more natural curved patterns.

The PPMs from all three methods are similar, showing large activation in the motor cortex and close to the central sulcus, and smaller activations in the somatosensory cortex. However, the activations detected by both of the proposed methods are slightly narrower and extend further along anatomical lines, indicating that the priors respect anatomical boundaries.

\vspace{-0.55cm}
\section{Discussion}
We have proposed two new Bayesian spatial priors for fMRI analysis that allow for a locally anisotropic spatial dependence over voxels. These priors were used to encode anatomical structure, resulting in anatomically-adaptive smoothing for fMRI data. The priors can be easily incorporated into the existing framework for Bayesian fMRI analysis, requiring minimal modifications.

While our presentation is centered on 2D priors, both of the proposed approaches can be extended to 3D without requiring significant modifications. The priors can also be improved by incorporating additional information from the structure tensor to, for example, use the UGL prior in areas with uniform intensity. These adaptive spatial models can also be applied in Bayesian frameworks for diffusion MRI data~\cite{gu2017bayesian}.

\clearpage
\newpage
\bibliographystyle{IEEEbib}
\bibliography{refs}

\begin{thebibliography}{10}

\bibitem{friman2003adaptive}
Ola Friman, Magnus Borga, Peter Lundberg, and Hans Knutsson,
\newblock ``{Adaptive analysis of fMRI data},''
\newblock {\em NeuroImage}, vol. 19, no. 3, pp. 837--845, 2003.

\bibitem{nandy2003novel}
Rajesh~R Nandy and Dietmar Cordes,
\newblock ``{Novel nonparametric approach to canonical correlation analysis
  with applications to low CNR functional MRI data},''
\newblock {\em Magnetic Resonance in Medicine}, vol. 50, no. 2, pp. 354--365,
  2003.

\bibitem{rydell2008bilateral}
Joakim Rydell, Hans Knutsson, and Magnus Borga,
\newblock ``{Bilateral filtering of fMRI data},''
\newblock {\em IEEE Journal of Selected Topics in Signal Processing}, vol. 2,
  no. 6, pp. 891--896, 2008.

\bibitem{eklund2011fast}
Anders Eklund, Mats Andersson, and Hans Knutsson,
\newblock ``{Fast random permutation tests enable objective evaluation of
  methods for single-subject fMRI analysis},''
\newblock {\em International journal of biomedical imaging}, vol. 2011, 2011.

\bibitem{behjat2015anatomically}
Hamid Behjat, Nora Leonardi, Leif S{\"o}rnmo, and Dimitri Van De~Ville,
\newblock ``{Anatomically-adapted graph wavelets for improved group-level fMRI
  activation mapping},''
\newblock {\em NeuroImage}, vol. 123, pp. 185--199, 2015.

\bibitem{zhuang2017family}
Xiaowei Zhuang, Zhengshi Yang, Tim Curran, Richard Byrd, Rajesh Nandy, and
  Dietmar Cordes,
\newblock ``{A family of locally constrained CCA models for detecting
  activation patterns in fMRI},''
\newblock {\em NeuroImage}, vol. 149, pp. 63--84, 2017.

\bibitem{lohmann2018lisa}
Gabriele Lohmann, Johannes Stelzer, Eric Lacosse, Vinod~J Kumar, Karsten
  Mueller, Esther Kuehn, Wolfgang Grodd, and Klaus Scheffler,
\newblock ``{LISA improves statistical analysis for fMRI},''
\newblock {\em Nature communications}, vol. 9, no. 1, pp. 4014, 2018.

\bibitem{penny2003variational}
Will Penny, Stefan Kiebel, and Karl Friston,
\newblock ``{Variational Bayesian inference for fMRI time series},''
\newblock {\em NeuroImage}, vol. 19, no. 3, pp. 727--741, 2003.

\bibitem{penny2005bayesian}
Will Penny and Guillaume Flandin,
\newblock ``{Bayesian analysis of fMRI data with spatial priors},''
\newblock in {\em Proceedings of the Joint Statistical Meeting (JSM). American
  Statistical Association}. Citeseer, 2005.

\bibitem{penny2007bayesian}
Will Penny, Guillaume Flandin, and Nelson Trujillo-Barreto,
\newblock ``{Bayesian comparison of spatially regularised general linear
  models},''
\newblock {\em Human brain mapping}, vol. 28, no. 4, pp. 275--293, 2007.

\bibitem{harrison2008diffusion}
Lee~M Harrison, W~Penny, Jean Daunizeau, and Karl~J Friston,
\newblock ``Diffusion-based spatial priors for functional magnetic resonance
  images,''
\newblock {\em Neuroimage}, vol. 41, no. 2, pp. 408--423, 2008.

\bibitem{harrison2008graph}
Lee~M Harrison, Will Penny, Guillaume Flandin, Christian~C Ruff, Nikolaus
  Weiskopf, and Karl~J Friston,
\newblock ``Graph-partitioned spatial priors for functional magnetic resonance
  images,''
\newblock {\em NeuroImage}, vol. 43, no. 4, pp. 694--707, 2008.

\bibitem{siden2017fast}
Per Sidén, Anders Eklund, David Bolin, and Mattias Villani,
\newblock ``{Fast Bayesian whole-brain fMRI analysis with spatial 3D priors},''
\newblock {\em NeuroImage}, vol. 146, pp. 211--225, 2017.

\bibitem{siden2019}
Per Sid{\'e}n, Finn Lindgren, David Bolin, Anders Eklund, and Mattias Villani,
\newblock ``Spatial {3D Matern} priors for fast whole-brain {fMRI} analysis,''
\newblock {\em arXiv preprint arXiv:1906.10591}, 2019.

\bibitem{chung1997spectral}
Fan~RK Chung,
\newblock {\em {Spectral graph theory}},
\newblock American Mathematical Society, 1997.

\bibitem{granlund1995signal}
Gösta~H Granlund and Hans Knutsson,
\newblock {\em {Signal processing for computer vision}},
\newblock Kluwer Academic Publishers, 1995.

\bibitem{knutsson2003what}
Hans Knutsson and Mats Andersson,
\newblock ``{What's so good about quadrature filters?},''
\newblock in {\em Proceedings 2003 International Conference on Image Processing
  (Cat. No. 03CH37429)}. IEEE, 2003, vol.~3, pp. III--61.

\bibitem{knutsson1989representing}
Hans Knutsson,
\newblock ``{Representing local structure using tensors},''
\newblock in {\em Scandinavian Conference on Image Analysis}, 1989, pp.
  244--251.

\bibitem{knutsson2011representing}
Hans Knutsson, Carl-Fredrik Westin, and Mats Andersson,
\newblock ``{Representing local structure using tensors II},''
\newblock in {\em Scandinavian Conference on Image Analysis}. Springer, 2011,
  pp. 545--556.

\bibitem{van2013wu}
David~C Van~Essen, Stephen~M Smith, Deanna~M Barch, Timothy~EJ Behrens, Essa
  Yacoub, Kamil Ugurbil, Wu-Minn~HCP Consortium, et~al.,
\newblock ``{The WU-Minn human connectome project: an overview},''
\newblock {\em Neuroimage}, vol. 80, pp. 62--79, 2013.

\bibitem{gu2017bayesian}
Xuan Gu, Per Sidén, Bertil Wegmann, Anders Eklund, Mattias Villani, and Hans
  Knutsson,
\newblock ``{Bayesian diffusion tensor estimation with spatial priors},''
\newblock in {\em International Conference on Computer Analysis of Images and
  Patterns}. Springer, 2017, pp. 372--383.

\end{thebibliography}

\end{document}